\documentclass[prl,aps,twocolumn,showpacs]{revtex4}
\usepackage{graphics}
\usepackage{graphicx}
\usepackage{amssymb}
\usepackage{amsmath}
\usepackage{bm}

\newcommand{\mytitle}[1]{
 \twocolumn[\hsize\textwidth\columnwidth\hsize
 \csname@twocolumnfalse\endcsname #1 \vspace{1mm}]}

\newcommand{\beq}{\begin{equation}}
\newcommand{\eeq}{\end{equation}}
\newcommand{\bea}{\begin{eqnarray}}
\newcommand{\eea}{\end{eqnarray}}

\begin{document}

\title{Non-equilibrium transport at a dissipative quantum phase transition}

\author{Chung-Hou Chung$^{1}$, Karyn Le Hur$^{2}$, Matthias Vojta$^{3}$, and Peter W\" olfle$^{4,5}$}
\affiliation{$^{1}$Electrophysics Department, National Chiao-Tung University, HsinChu,
Taiwan \\
$^{2}$ Department of Physics and Applied Physics, Yale University, New
Haven, CT, USA \\
$^{3}$Institut f\"ur Theoretische Physik, Universit\"at zu K\"oln, 50937 K\"oln, Germany\\
\mbox{$^{4}$Institut f\"ur Theorie der Kondensierten Materie, Universit\"at
Karlsruhe, 76128 Karlsruhe, Germany} \\
$^{5}$ INT, Forschungszentrum Karlsruhe, 76021 Karlsruhe, Germany} 
\date{\today}

\begin{abstract}
We investigate the non-equilibrium transport near a quantum phase transition
in a generic and relatively simple model, the dissipative resonant
level model, that has many applications for nanosystems.
We formulate a rigorous mapping and apply a controlled frequency-dependent
renormalization group approach to compute the non-equilibrium current
in the presence of a finite bias voltage $V$ and a finite temperature $T$.
For $V\rightarrow 0$, we find that the conductance has its well-known
equilibrium form, while it displays a distinct non-equilibrium profile
at finite voltage.
\end{abstract}

\maketitle


In recent years, quantum phase transitions (QPTs) \cite{sachdevQPT,Steve} have
attracted much attention at the nanoscale \cite{lehur1,lehur2,zarand,Markus,matveev,Zarand2}.
 A finite bias voltage applied across a nanosystem is
expected to smear out the equilibrium transition, but the current-induced decoherence
might act quite differently as compared to thermal decoherence at finite temperature $T$,
resulting in exotic behavior near the transition. Non-equilibrium effects at a quantum phase transition appear as an emerging field both in experimental and theoretical condensed matter
physics  \cite{Feldman,mitra,Goldhaber,Si}. Quantum impurity systems out of equilibrium are also extensively studied theoretically \cite{Andrei}. In this Letter, we aim to answer several fundamental questions related to the non-equilibrium transport in quantum dot settings,
such as what is the scaling behavior of the conductance at
zero temperature and finite bias voltage near the transition. 

For this purpose, we employ a typical nano-model comprising a dissipative
resonant level (quantum dot).
In this model, the QPT separating the conducting and insulating phase
for the level is solely driven by dissipation, which can be modeled by a
bosonic bath. Dissipation-driven QPTs have been
addressed theoretically and experimentally in various systems, such as:
Josephson junction arrays \cite{Josephson}, superconducting thin films \cite%
{Kapitulnik}, superconducting qubits \cite{Rimberg} and biological systems
\cite{McKenzie}. 

Our Hamiltonian takes the precise form:
\begin{eqnarray}
H &=& \sum_{k,i=1,2} (\epsilon(k)-\mu_i) c^{\dagger}_{k i}c_{k i} + t_{i}
c^{\dagger}_{ki} d + h.c. \\
&+& \sum_{r} \lambda_{r} (d^{\dagger}d-1/2) (b_{r} + b^{\dagger}_{r}) +
\sum_{r} \omega_{r} b^{\dagger}_{r} b_{r},  \nonumber
\end{eqnarray}
where $t_{i}$ is the (real) hopping amplitude between the
lead $i$ and the quantum dot, $c_{ki}$ and $d$ are electron operators for the
(Fermi-liquid type) leads and the quantum dot, respectively.
$\mu_i = \pm V/2$ is the chemical potential applied on the lead $i$,
while the dot level is at zero chemical potential.
\begin{figure}[h]
\begin{center}
\vspace{-0.2cm} 
\includegraphics[width=6cm]{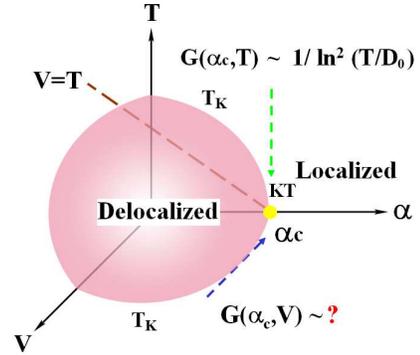}
\end{center}
\par
\vskip -0.8cm
\label{3dphase}
\caption{Schematic phase diagram as a function of
temperature $T$, dissipation strength $\protect\alpha$, and bias voltage $V$.
}
\vskip -0.4cm
\end{figure}
To simplify the discussion, we assume
that the electron spins have been polarized through the application of a
strong magnetic field. Here, $b_{\alpha}$ are the boson operators of the
dissipative bath, that is governed by an ohmic spectral density 
\cite{lehur2}: $\mathit{J}(\omega) = \sum_{r} \lambda_{r}^2
\delta(\omega-\omega_{r}) = \alpha \omega$. {\it We use units in which $\hbar=1$ and electric charge $e=1$}.

In equilibrium ($V\!=\!0$), such a dissipative system comprising several leads
maps onto the anisotropic one-channel Kondo model; we denote by $t$ the
hopping amplitude between the (effective) lead
and the level. We introduce the dimensionless transverse Kondo coupling $g_{\perp}^{(e)}$
which is proportional to $t$ (the exact prefactor is given in Refs. \cite{lehur2,Markus,matveev}) and the longitudinal coupling $g_{z}^{(e)}\propto 1-\sqrt{\alpha }$ \cite{lehur2,Markus,matveev}.
The model exhibits a Kosterlitz-Thouless (KT) QPT from a delocalized
(Kondo screened) phase for $g_{\perp}^{(e)}+g_{z}^{(e)}>0$,
with a large conductance, $G\approx 1/h$ ($e=1$ and $h=2\pi\hbar =2\pi$),
to a localized (local moment) phase for $g_{\perp}^{(e)}+g_{z}^{(e)}\leq 0$,
with a small conductance, as the dissipation strength is increased (see Fig. 1).
For $g_{\perp}^{(e)}\rightarrow 0$, the KT transition occurs at $\alpha _{c}=1$.
As $\alpha\to\alpha_c$, the Kondo temperature $T_{K}$
obeys $\ln T_{K}\propto 1/(\alpha-\alpha_c)$ \cite{lehur1}.

In equilibrium, the scaling functions $g_{\perp }^{(e)}(T)$ and $g_{z}^{(e)}(T)$ at the quantum critical point
are obtained via the renormalization-group (RG) equations of the anisotropic Kondo
model: $g_{\perp ,cr}^{(e)}(T)=-g_{z,cr}^{(e)}(T)=(2\ln \left(T_D/{T}\right) )^{-1}$;
hereafter, we introduce the energy scale $T_D=D_{0}e^{1/(2g_{\perp })}$, with $D_{0}$ being the
ultraviolet cutoff and we set the Boltzmann constant $k_B=1$. Having in mind a quantum dot at resonance, $D_{0}=\min
(\delta \epsilon ,\omega _{c})$, with $\delta \epsilon $ being the level
spacing on the dot and $\omega _{c}$ the cut-off of the bosonic bath; $D_{0}$
is of the order of a few Kelvins. At the KT quantum phase transition, the conductance drops
abruptly \cite{zarand}:
\begin{equation}
G_{eq}(\alpha _{c},T\ll D_0)\propto \left[ g_{\perp ,cr}^{(e)}(T)\right]
^{2}\propto \frac{1}{\ln ^{2}(T/T_D)}.  \label{GTeq}
\end{equation}
Below, we analyze the nonequilibrium $(V\neq 0$) transport at the phase
transition and in the localized phase.

First, we envision a non-equilibrium mapping revealing that the leads are
controlled by distinct chemical potentials. Through similar bosonization and
refermionization procedures as in equilibrium 
\cite{lehur1,lehur2,Markus,matveev}, our model is mapped onto an anisotropic
Kondo model with the effective (Fermi-liquid) left ($L$) and right lead ($R$
) \cite{mapping}:
\begin{eqnarray}
{H}_{K} &=&\sum_{k,\gamma =L,R,\sigma =\uparrow ,\downarrow }[\epsilon
_{k}-\mu _{\gamma }]c_{k\gamma \sigma }^{\dagger }c_{k\gamma \sigma } \\
&+&(J_{\perp }^{1}s_{LR}^{+}S^{-}+J_{\perp
}^{2}s_{RL}^{+}S^{-}+h.c.)+\sum_{\gamma =L,R}J_{z}s_{\gamma \gamma
}^{z}S^{z},  \nonumber
\end{eqnarray}
where $c_{kL(R)\sigma }^{\dagger }$ is the electron operator of the
effective lead $L(R)$, with $\sigma$ the spin quantum
number, $S^{+}=d^{\dagger }$, $S^{-}=d$, and $S^{z}=Q-1/2$ where $%
Q=d^{\dagger }d$ describes the charge occupancy of the level. Additionally, $%
s_{\gamma \beta }^{\pm }=\sum_{\alpha ,\delta ,k,k^{\prime }}1/2c_{k\gamma
\alpha }^{\dagger }\mathbf{\sigma }_{\alpha \delta }^{\pm }c_{k^{\prime
}\beta \delta }$ are the spin-flip operators between the effective leads $%
\gamma $ and $\beta $, $J_{\perp }^{1(2)}\propto {t_{1(2)}}$ embody the
transverse Kondo couplings, $J_{z}\propto 1/2(1-{1}/\sqrt{2\alpha ^{\ast }})$%
, and $\mu _{\gamma }=\pm \frac{V}{2}\sqrt{1/(2\alpha ^{\ast })}$, where $%
1/\alpha ^{\ast }=1+\alpha $. It should be noted that this mapping is exact near the phase
transition where $\alpha \rightarrow 1$ or $\alpha ^{\ast }\rightarrow 1/2$,
and thus $\mu _{\gamma }=\pm V/2$.

From the
mapping, $N_1-N_2 = (N_{L}-N_{R})$, where 
$N_i=\sum_{ki} c^{\dagger}_{ki} c_{ki}$ represents the charge in lead $i=1,2$, whereas $N_{\gamma}= \sum_{k} c^{\dagger}_{k\gamma\sigma} c_{k\gamma\sigma}
$ represents the charge in the effective lead $\gamma=L,R$. This allows us to check that the averaged currents within the
Keldysh formalism are the same in the original and in the effective Kondo
model. Thus, the current $I$ can be computed from the Kondo model.

The poor-man scaling equations of Anderson are generalized to
nonequilibrium RG equations by including the frequency dependence of the
Kondo couplings and the decoherence due to the steady-state current at
finite bias voltage \cite{noneqRG}. For the sake of clarity, we assume that
the resonant level (quantum dot) is symmetrically coupled to the right and
to the left lead, $t_{1}=t_{2}$. The dimensionless Kondo couplings then have
the extra symmetry ($\omega $ is the frequency): $g_{\perp (z)}(\omega
)=g_{\perp (z)}(-\omega )$ where $g_{\perp (z)}=N(0)J_{\perp (z)}$ with $N(0)
$ being the density of states per spin of the conduction electrons. We
obtain \cite{noneqRG}:
\begin{eqnarray}
\frac{\partial g_{z}(\omega )}{\partial \ln D} &=&-\sum_{\beta =-1,1}\left[
g_{\perp }\left( \frac{\beta V}{2}\right) \right] ^{2}\Theta _{\omega +
\frac{\beta V}{2}}  \label{gpergz} \\
\frac{\partial g_{\perp }(\omega )}{\partial \ln D} &=&-\!\!\!\sum_{\beta
=-1,1}\!\!\!g_{\perp }\left( \frac{\beta V}{2}\right) g_{z}\left( \frac{\beta V}{2}\right) \Theta _{\omega +\frac{\beta V}{2}},  \nonumber
\end{eqnarray}
where $\Theta _{\omega }=\Theta (D-|\omega +\mathit{i}\Gamma |)$, $D<D_{0}$
is the running cutoff, and $\Gamma $ is the decoherence (dephasing) rate at
finite bias which cuts off the RG flow \cite{noneqRG}. The configurations of
the system out of equilibrium are not true eigenstates, but acquire a finite
lifetime. The spectral function of the fermion on the level is peaked
at $\omega =\pm V/2$, and therefore we have $g_{\perp (z)}(\omega )\approx
g_{\perp (z)}(\pm V/2)$ on the right hand side of Eq. (\ref{gpergz}).
Other Kondo couplings are not generated.

In the Kondo model, $\Gamma$ corresponds to the relaxation rate due to
spin flip processes (which are charge flips in the original model).
From Ref.~\cite{noneqRG}, we identify:
\begin{eqnarray}  \label{gamma}
\Gamma = \frac{\pi}{4} \sum_{\gamma,\gamma^{\prime},\sigma} \int \!\!
d\omega \Bigl[ n_{\sigma} g_{z}^2(\omega)
f_{\omega-\mu_{\gamma}}(1-f_{\omega-\mu_{\gamma ^{\prime}}}) && \\
+ n_{\sigma} g_{\perp}^2(\omega)
f_{\omega-\mu_{\gamma}}(1-f_{\omega-\mu_{\gamma^{\prime}}})\Bigr], &&
\nonumber
\end{eqnarray}
where $f_{\omega}$ is the Fermi function. Here, $\gamma=\gamma^{\prime}$ for
the $g_z^2(\omega)$ terms while $\gamma \neq \gamma^{\prime}$ for the
$g_{\perp}^2(\omega)$ terms with $\gamma$, $\gamma^{\prime}$ being $L$ or $R$.
We have introduced the occupation numbers $n_\sigma$ for up and down
spins satisfying $n_{\uparrow} +n_{\downarrow}=1$ and $S_z=1/2(n_{%
\uparrow}-n_{\downarrow})$. In the delocalized phase, we get $n_{\uparrow} =
n_{\downarrow} = 1/2$ in agreement with the quantum Boltzmann equation \cite%
{noneqRG}; at the phase transition we can use that $g_{\perp}(\omega)=-g_z(%
\omega)$ and that $\sum_{\sigma} n_{\sigma} =1$, and finally in the
localized phase $g_{\perp}\le -g_z$, $n_{\sigma}$ satisfies $|S_z|
\rightarrow 1/2$ \cite{lehur1,lehur2,Markus,matveev}.

Following the scheme of Ref. \cite{noneqRG}, we solve Eqs. (\ref{gpergz})
and (\ref{gamma}) self-consistently. First, we compute $g_{\perp(z)}(\omega
=\pm V/2)$ for a given cutoff $D$. Second, we substitute the solutions back
into the RG equations to get the general solutions for $g_{\perp(z)}(\omega)$
at finite $V$, and then extract the solutions in the limit $D\to 0$. When
the cutoff $D$ is lowered, the RG flows are not cutoff by $V$ but continue
to flow for $\Gamma < D < V$ until they are stopped for $D<\Gamma$.

\begin{figure}[t]
\begin{center}
\includegraphics[width=6.6cm]{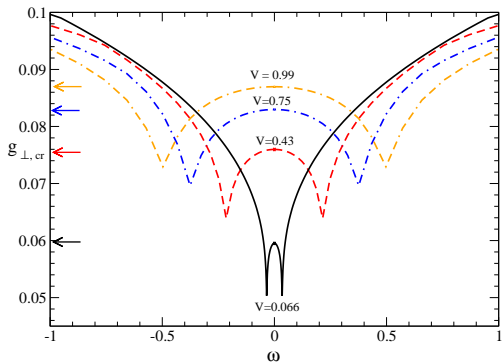}
\end{center}
\par
\vspace{-0.5cm} 
\caption{$g_{\perp,cr}(\protect\omega)= -g_{z,cr}(\protect\omega)$ at $\alpha=\alpha_c$; the bare couplings are $g_{\perp}=-g_z = 0.1$. We have set
$V = 0.066 D_0$, $0.43 D_0$, $0.75 D_0$ and $0.99 D_0$ where $D_0 = 1$ for all the figures. The
arrows give the values of $g_{\perp,cr}(\protect\omega = 0)$ at these bias
voltages. }
\label{gpergzfig}
\end{figure}

At the KT transition, we both numerically and analytically solve Eqs. (\ref%
{gpergz}) and (\ref{gamma}) (in the limit of $D\to 0$):
\begin{eqnarray}  \label{gsolution}
g_{\perp,cr}(\omega) &=& \sum_{\beta} \Theta (|\omega-\beta V/2|-V) \frac{1}{%
4 \ln\left[\frac{T_D}{|\omega-\beta V/2|}\right]} \\
&+& \Theta (V-|\omega-\beta V/2|) \times  \nonumber \\
&& \left[ \frac{1}{\ln[T_D^{2}/ V \max(|\omega-\beta V/2|,\Gamma)]}
- \frac{1}{4 \ln\frac{T_D}{V}}\right].  \nonumber
\end{eqnarray}
The solutions at the transition (denoted $g_{\perp,cr}$ and $g_{z,cr}$) are shown in Fig. \ref{gpergzfig}. Since $g_{\perp,cr}(\omega)$
decreases under the RG scheme, the effect of the
decoherence leads to minima; the couplings are 
severely suppressed at the points $\omega=\pm \frac{V}{2}$. We
also check that $g_{\perp,cr}(\omega) = - g_{z,cr}(\omega)$.

\begin{figure}[t]
\begin{center}
\includegraphics[width=6.6cm]{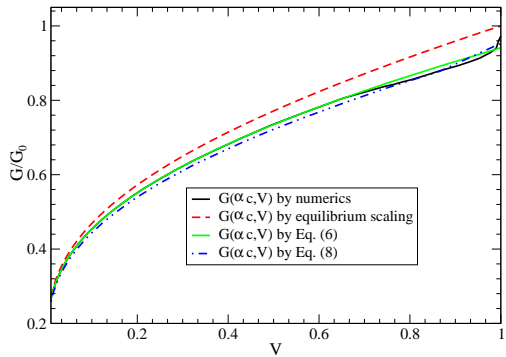}
\end{center}
\par
\vspace{-0.5cm}
\caption{Nonequilibrium conductance at the KT transition. $G_0$ is the
equilibrium conductance at the transition for $T=D_0$: $G_0 = G_{eq}(\protect%
\alpha_c, T=D_0) = 0.005 \protect\pi$ with the bare couplings $%
g_{\perp} = -g_z = 0.1$. Again, we set the charge $e=\hbar=1$. }
\label{Icr}
\end{figure}

From the Keldysh calculation up to second order in the tunneling amplitudes,
the current reads:
\begin{eqnarray}  \label{noneqI}
I = \frac{ \pi}{8} \int d\omega \Big[\sum_\sigma 4g_{\perp}(\omega)^2
n_{\sigma}\times \\
f_{\omega-\mu_{L}} (1-f_{\omega-\mu_{R}})\Big] - (L \leftrightarrow R).
\nonumber
\end{eqnarray}
At $T=0$, it simplifies as $I = \frac{\pi}{2} \int_{-V/2}^{V/2}
\! d\omega  g_{\perp}^{2}(\omega)$. Then, we numerically evaluate the
nonequilibrium current. The conductance is obtained from $G(V) = dI/dV$. The
$T=0$ results at the KT transition are shown in Fig. \ref{Icr}. First, it is
instructive to compare the non-equilibrium current at the transition to the
approximate expression:
\begin{eqnarray}  \label{IVapprox}
I(\alpha_c,V) &\approx & \frac{\pi V}{2} \left( \frac{\pi}{4} \left[%
g_{\perp,cr}(\omega = 0)\right]^2 \right) \\
& +& \frac{\pi V}{2} \left( (1-\frac{\pi}{4}) \left[g_{\perp,cr}(\omega
= V/2)\right]^2 \right),  \nonumber
\end{eqnarray}
where $g_{\perp,cr}(\omega = 0) \approx 2 \left(\frac{1}{\ln(2T_D^2/V^2)} - \frac{1}{4\ln(T_D/V)}\right)$, and $g_{\perp,cr}(\omega
= V/2) \approx 1/\ln\left(\frac{T_D^2}{\Gamma V}\right)$. We have treated $%
g_{\perp,cr}(\omega)^2$ within the interval $-V/2<\omega<V/2$ as a
semi-ellipse.

As demonstrated in Fig. \ref{Icr}, the conductance $G(V)$ obtained via 
the approximation in Eq. (\ref{IVapprox}) fits very well with that obtained
numerically over a whole range of $0<V<D_0$. In the low-bias $V\to 0$
(equilibrium) limit, since $g_{\perp,cr}(\omega = 0)\approx
g_{\perp,cr}^{(e)}(T=V)\ll 1$, we have $I(\alpha_c,V)\approx \frac{\pi V}{2}
\left( g_{\perp,cr}^{(e)}(T= V)\right)^2$; therefore the scaling of 
$G(\alpha_c, V)$ is reminiscent of the equilibrium expression in 
Eq. (\ref{GTeq}), $G(\alpha_c,V) \approx \frac{\pi}{2} \left(g_{%
\perp,cr}^{(e)}(T=V)\right)^2 = \frac{\pi}{8}\frac{1}{\ln^2(T_D/V)}$. This agreement between equilibrium and nonequilibrium conductance
at low $V$ persists up to a crossover scale $V\approx 0.01D_0$ (determined for the parameters used
in Fig. 3). At larger biases, the conductance shows a unique nonequilibrium profile; see Eq. (\ref{IVapprox}). 
We find an excellent agreement of the nonequilibrium conductance obtained by three
different ways --- pure numerics, analytical solution Eq. (\ref{gsolution})
and the approximation in Eq. (\ref{IVapprox}).

The distinct nonequilibrium scaling behavior seen here is in fact closely
tied to the non-trivial (non-linear) $V$ dependence of the decoherence rate $%
\Gamma (V)$. In particular, at the KT transition, we find that $\Gamma \sim
\frac{1}{2}I$ is a highly non-linear function in $V$, resulting in the
deviation of the nonequilibrium scaling from that in equilibrium. For large bias voltages $V\rightarrow D_{0}$, since $g_{\perp,cr}(\omega )$ approaches its bare value $g_{\perp}$, the nonequilibrium conductance increases rapidly and reaches $G(\alpha _{c},V)\approx G_{0}=\frac{\pi }{%
2}g_{\perp}^{2}$. The nonequilibrium conductance is 
smaller than the equilibrium one, $G(\alpha _{c},V)<G_{eq}(\alpha _{c},T=V)$, since $g_{\perp
}(\omega =\pm V/2)<g_{\perp }(\omega =0)$. Additionally, in the delocalized phase for 
$V\gg T_{K}>0$, the RG flow of $g_{\perp}$ is suppressed by the
decoherence rate, and $G\propto 1/\ln ^{2}(V/T_{K})$ \cite%
{noneqRG}.

In the localized phase, the equilibrium RG equations
of the effective Kondo model can be solved analytically, resulting in $G_{loc}^{(e)}(T)= \frac{\pi}{%
2} \left(g_{\perp,loc}^{(e)}(T) \right)^2$, where 
\begin{equation}
g_{\perp,loc}^{(e)}(T) = \frac{2cg_{\perp} (c+|g_z|)}{(c+|g_z|)^2 - g_{\perp}^2
(\frac{T}{D_0})^{4c}} \left(\frac{T}{D_0}\right)^{2c},
\end{equation}
with $c = \sqrt{g_z^2 -
g_{\perp}^2}$. We introduce the energy scale 
${T}^{\ast}=D_0 e^{-\pi/\sqrt{g_z^2-g_{\perp}^2}}$ (which vanishes at the KT transition) such that $g_{\perp,loc}^{(e)}(T)\propto ({T}/T^{\ast})^{2c}$  for $T\to 0$, leading to 
$G_{loc}^{(e)}(T)\propto ({T}/T^{\ast})^{4c}$.
\begin{figure}[h]
\begin{center}
\vspace{0.3cm} 
\includegraphics[width=8.25cm]{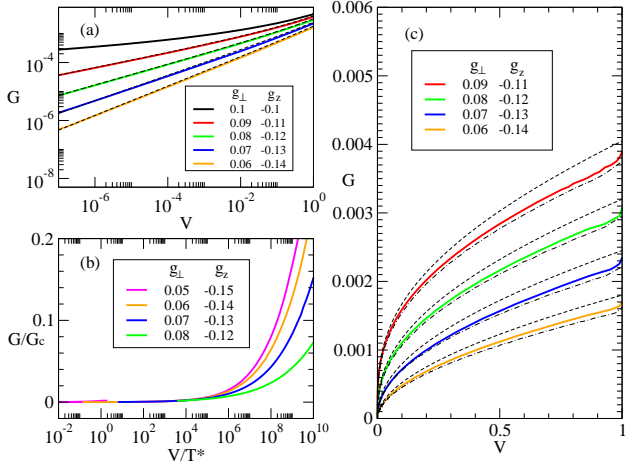}
\end{center}
\par
\vspace{-0.6cm}
\caption{
Conductance in the localized phase (in units of $\protect\pi/\hbar$%
). (a) $G(V)$ at low bias follows the equilibrium scaling
(dashed lines). (b) The conductance $G(V)/G_c$ is a function of $V/{T}^{\ast}$ where we have defined $G_c=G(\protect\alpha_c, V)$ and ${T}^{\ast}=D_0 e^{-\protect\pi/%
\protect\sqrt{g_z^2-g_{\perp}^2}}$. (c) At large bias voltages $V$, the nonequilibrium
conductance $G(V)$ (solid lines) is distinct from the equilibrium form
(dashed lines). The dot-dashed lines stem from an analytical approximation
via Eq. \protect\eqref{IVapprox}.}
\label{Gv}
\end{figure}
For very small bias voltages $V\to 0$, we find that the conductance reduces to the
equilibrium scaling: $G(V)\to G_{loc}^{(e)}(T=V) \propto ({V}/{{T}^{\ast}})^{4c}$ (see
Fig. \ref{Gv} (a) and Fig. 4 (b)). For $g_{\perp,loc}\ll |g_{z,loc}|$ and $\alpha^{\ast}\to 1/2$, we
get that the exponent $4c \approx 2\alpha^{\ast} - 1$, in perfect agreement
with that obtained in equilibrium at low temperatures: $G(T)\propto
T^{2\alpha^{\ast} - 1}$ \cite{zarand}. At higher bias voltages $0.01D_0<V< D_0
$, the conductance now follows a unique nonequilibrium form (consult Fig. \ref%
{Gv}(c)). 

\begin{figure}[!tbp]
\begin{center}
\vspace{0.5cm} 
\includegraphics[width=8.25cm]{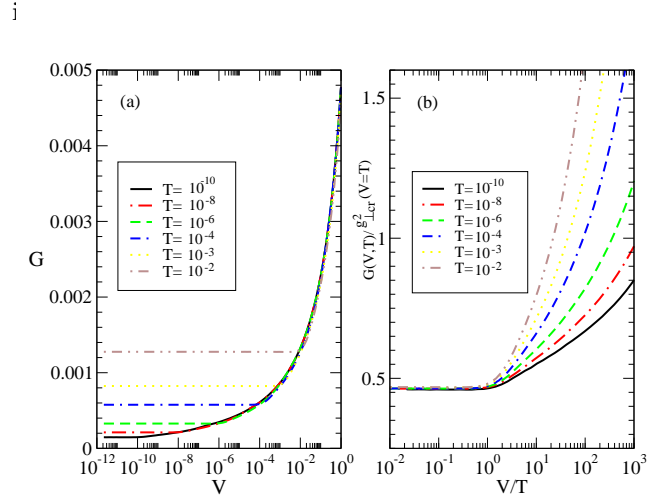}
\end{center}
\par
\vspace{-0.50cm} 
\caption{Scaling of the conductance at the KT transition (same unit as
in Fig. \protect\ref{Icr}). (a). For $V>T$, the conductance follows the
nonequilibrium scaling $G(\protect\alpha_c,V)$. (b). For $V<T$, now the
conductance follows the equilibrium scaling $G(\protect\alpha_c,T)$.}
\label{GvT}
\end{figure}

We have also analyzed the finite temperature profile of the nonequilibrium
conductance at the transition. We distinguish two different behaviors. For $V>T
$, the conductance $G(V,T)$ follows the nonequilibrium form at $%
T=0$ (see Fig. \ref{GvT}(a)), while for $V<T$ it follows the ($V=0$)
finite-temperature expression (see Fig. \ref{GvT}(b)). These two 
scaling behaviors have a crossover at $V=T$.

In summary, we have investigated the nonequilibrium transport at a
QPT using a standard nano-model, the dissipative resonant level.
We have used an exact mapping and applied a controlled frequency-dependent
renormalization group approach to compute the current.
For $V\rightarrow 0$, the conductance $G$ follows the equilibrium behavior; by increasing $V$, the
frequency-dependence of the couplings begins to play an important role and
therefore we systematically find very distinct scalings. We have also
analyzed the finite temperature profile of $G(V,T)$ at the transition and
identified two distinct behaviors at $V>T$ and $V<T$.
Finally, our results have a direct experimental relevance for dissipative two-level
systems.

We are grateful to D. Goldhaber-Gordon and G. Zarand for stimulating
discussions, and to R.T. Chang and K.V.P. Lata for technical support. We also
acknowledge the generous support from the NSC grant
No.95-2112-M-009-049-MY3, the MOE-ATU program, the NCTS of Taiwan, R.O.C. (C.H.C.),
the Department of Energy in USA under the
contract DE-FG02-08ER46541 (K.L.H.), the DFG via SFB 608 and SFB/TR-12 (M.V.), and
the DFG-Center for Functional Nanostructures, C.F.N. (P.W.). C.H.C. has also benefitted from the
visiting programs of KITP, ICTP, and MPI-PKS. 



\begin{references}
\bibitem{mapping}
In the delocalized phase, it would be more judicious to redefine $\Psi \to (t_1 c_1+ t_2 c_2)/\sqrt{t_1^2+t_2^2}$, and then bosonize and refermionize to get the fermions with spins \cite{lehur2}, $\Psi_{\sigma}$.
\end{references}
\end{document}